\documentclass[prl,twocolumn,superscriptaddress]{revtex4}
\usepackage{graphicx}

\begin{document}

\title{Electron Emission from Diamondoids: A Diffusion Quantum Monte
  Carlo Study}

\author{N.~D.~Drummond}

\affiliation{TCM Group, Cavendish Laboratory, University of Cambridge,
Cambridge CB3 0HE, United Kingdom}

\author{A.~J.~Williamson}

\email{williamson10@llnl.gov}

\affiliation{Lawrence Livermore National Laboratory, Livermore, CA
94550}

\author{R.~J.~Needs}

\affiliation{TCM Group, Cavendish Laboratory, University of Cambridge,
Cambridge CB3 0HE, United Kingdom}

\author{G.~Galli}

\affiliation{Lawrence Livermore National Laboratory, Livermore, CA
94550}

\begin{abstract}
  We present density-functional theory (DFT) and quantum Monte Carlo
  (QMC) calculations designed to resolve experimental and theoretical
  controversies over the optical properties of H-terminated C
  nanoparticles (diamondoids). The QMC results follow the trends of
  well-converged plane-wave DFT calculations for the size dependence
  of the optical gap, but they predict gaps that are 1--2\,eV higher.
  They confirm that quantum confinement effects disappear in
  diamondoids larger than 1\,nm, which have gaps below that of bulk
  diamond.  Our QMC calculations predict a small exciton binding
  energy and a negative electron affinity (NEA) for diamondoids up to
  1\,nm, resulting from the delocalized nature of the lowest
  unoccupied molecular orbital. The NEA suggests a range of possible
  applications of diamondoids as low-voltage electron emitters.
\end{abstract}

\pacs{73.22.-f, 02.70.Ss}

\maketitle

Designing the nanoscale building blocks for future nanotechnological
devices is one of the most rapidly evolving fields of materials
science.  In the area of nanoscale optoelectronic materials, attention
was originally focused on semiconductor nanoparticles constructed from
Si~\cite{wilson93} or Ge~\cite{ribeiro98}, as these nanoparticles can
easily be integrated with existing Si fabrication techniques.
Recently, however, it was discovered that large quantities of
high-quality, H-terminated C nanoparticles can be isolated from
petroleum and separated into mono-disperse samples.\cite{dahl2003}
These \textit{diamondoids} are expected to have several
technologically useful optoelectronic properties.  In particular, the
effect of quantum confinement is expected to push diamondoid optical
gaps into the UV range, enabling a unique set of sensing applications.
Additionally, it has been demonstrated\cite{himpsel79,rutter98} that
H-terminated diamond surfaces exhibit negative electron affinities
(NEA's), suggesting that diamondoids will also have NEA's. This opens
up the possibility of coating surfaces with diamondoids to produce new
electron-emission devices.

The breakthrough in isolating diamondoids has sparked interest in
measuring~\cite{chang99,raty2003,hosokawa2004} and calculating their
structural and optoelectronic
properties~\cite{raty2003,barnard2003,lee2003,park2004,mcintosh2004}.
However, such studies have proved to be challenging, and have produced
several controversial results.  In 1999, the results of \textit{X-ray
absorption near-edge structure} (XANES) studies of diamond films
produced by chemical vapor deposition (CVD)\cite{chang99} were used to
infer the evolution of the nanoparticle gap with size.  This showed a
persistence of quantum confinement effects up to 27\,nm, much longer
than in Si or Ge nanoparticles, where quantum confinement effects
disappear above 5--7\,nm. In contrast, recent \textit{near-edge
absorption fine structure} (NEXAFS) studies of diamondoids prepared by
hot-filament CVD\cite{tang2003} and high-explosive detonation
waves\cite{raty2003} showed that quantum confinement effects disappear
in particles larger than 4\,nm.

Theoretical models for the optoelectronic properties of diamondoids
have also produced contradictory results.  Raty \textit{et
al.}~\cite{raty2003} studied the size dependence of the optical gap
using both time-independent and time-dependent density-functional
theory (DFT) calculations, and concluded that quantum confinement
effects disappear in nanoparticles larger than 1\,nm.  Raty \textit{et
al.}~\cite{raty2003} also predicted that the gaps of diamondoids with
sizes between 1 and 1.5\,nm are {\em below} the gap of bulk diamond.
This is strikingly different to the behavior of H-terminated Si and Ge
nanoparticles, for which the gaps are always above the bulk band
gap.\cite{williamson2002} In contrast, recent DFT calculations by
McIntosh \textit{et al.}~\cite{mcintosh2004} for the same particles
predict optical gaps 2\,eV above the gap of bulk diamond for particles
ranging in size from 0.5 to 2\,nm.

In this Letter we resolve these controversies by performing both DFT
and highly accurate quantum Monte Carlo (QMC) calculations to predict
the size dependence of the optical gap and the electron affinity of
diamondoids.   We have studied two classes of nanoparticles: (i)
diamondoids constructed from adamantane cages, {\em adamantane},
C$_{10}$H$_{16}$, {\em diamantane}, C$_{14}$H$_{20}$, and {\em
pentamantane}, C$_{26}$H$_{32}$; and (ii) H-terminated, spherical,
diamond-structure nanoparticles, C$_{29}$H$_{36}$, C$_{66}$H$_{64}$,
and C$_{87}$H$_{76}$.  The nanoparticle sizes and symmetries are given
in Table~\ref{gap_table}.  Because diamondoids can be extracted in
large quantities from petroleum and highly purified using
high-pressure liquid chromatography, we anticipate that actual
experimental samples will consist largely of the high-symmetry
structures studied in this work.  This is not the case for Si and Ge
nanoparticles, where limitations in current synthesis techniques
prevent the routine production of high-symmetry nanoparticles.

\begin{table}
\begin{tabular}{lccr@{.}lr@{.}lcr@{.}l}
\hline \hline

Cluster & Sym  & Size & \multicolumn{2}{c}{DFT gap} &
\multicolumn{2}{c}{DMC gap} & EA & \multicolumn{2}{c}{IP} \\

\hline

C$_{10}$H$_{16}$ & $T_d$    & $0.50$   & \hspace{0.2cm} $5$&$77$  &  ~
$7$&$61(2)$ & $-0.13(2)$ & $10$&$15(3)$ \\

C$_{14}$H$_{20}$ & $D_{3d}$ & $0.69$   & $5$&$41$      & $7$&$32(3)$ \\

C$_{26}$H$_{32}$ & $T_d$ &   $0.74$   & $5$&$03$      & $7$&$04(6)$ \\

C$_{29}$H$_{36}$ & $T_d$ &   $0.76$   & $4$&$90$      & $6$&$67(6)$ &
$-0.29(6)$ & $7$&$63(5)$ \\

C$_{66}$H$_{64}$ & $T_d$ &   $1.00$   & $4$&$37$      & $5$&$09(18)$ \\

C$_{87}$H$_{76}$ & $T_d$  &  $1.14$   & $3$&$94$       \\

Diamond          &   &       & $4$&$23$\cite{raty2003} &
$5$&$6(2)$\cite{towler2000}\\

\hline \hline
\end{tabular}
\caption{\label{gap_table} Point-group symmetry, size (in nm), DFT and
  DMC optical gaps, DMC electron affinities (EA), and DMC ionization
  potentials (IP) (all in eV) of different diamondoids. }
\end{table}

Our DFT calculations were performed using both plane-wave and Gaussian
basis sets.  The PBE~\cite{pbe} functional was used in all the DFT
calculations, as this has been shown to reproduce accurately the
structural and electronic properties of diamond.\cite{raty2003} The
plane-wave calculations were performed with the GP~\cite{GP},
QBox\cite{qbox}, and \textsc{abinit}~\cite{abinit} codes.  A
plane-wave cutoff of 55\,Ry and a cubic simulation cell of side-length
30\,\AA~were used for all calculations.  Norm-conserving
Troullier-Martins (TM) pseudopotentials were used to represent the
ionic cores.  The all-electron Gaussian calculations were performed
with the \textsc{G03} code.\cite{gaussian}

The initial structures of C$_{10}$H$_{16}$, C$_{14}$H$_{20}$, and
C$_{26}$H$_{32}$ were taken from Ref.~\cite{dahl2003}, while those of
C$_{29}$H$_{36}$, C$_{66}$H$_{64}$, and C$_{87}$H$_{76}$ were
constructed from the bulk-diamond structure, and the C-H bond lengths
were set to the experimental value in CH$_4$.  All the structures were
fully relaxed to their lowest-energy configurations within DFT before
their electronic and optical properties were calculated.  For
C$_{10}$H$_{16}$, C$_{14}$H$_{20}$, and C$_{26}$H$_{32}$, DFT-PBE
calculations predict small relaxations in the bond lengths and angles,
similar to Ref.~\cite{mcintosh2004} The C-C and C-H bond lengths are
1.55 and 1.10\,\AA, respectively, essentially identical to the
bulk-diamond and methane bond lengths.  The bond angles vary from
$108.5^\circ$ to $110.5^\circ$.  For the larger, spherical particles
(C$_{29}$H$_{36}$, C$_{66}$H$_{64}$, and C$_{87}$H$_{76}$), DFT-PBE
calculations predict a 3--4\% increase in the C-C bond lengths, while
the C-H bonds remain at 1.10\,\AA.  The bond angles vary from
$102^\circ$ to $113^\circ$.

\begin{figure}[t]
  \includegraphics[width=\linewidth,clip=true]{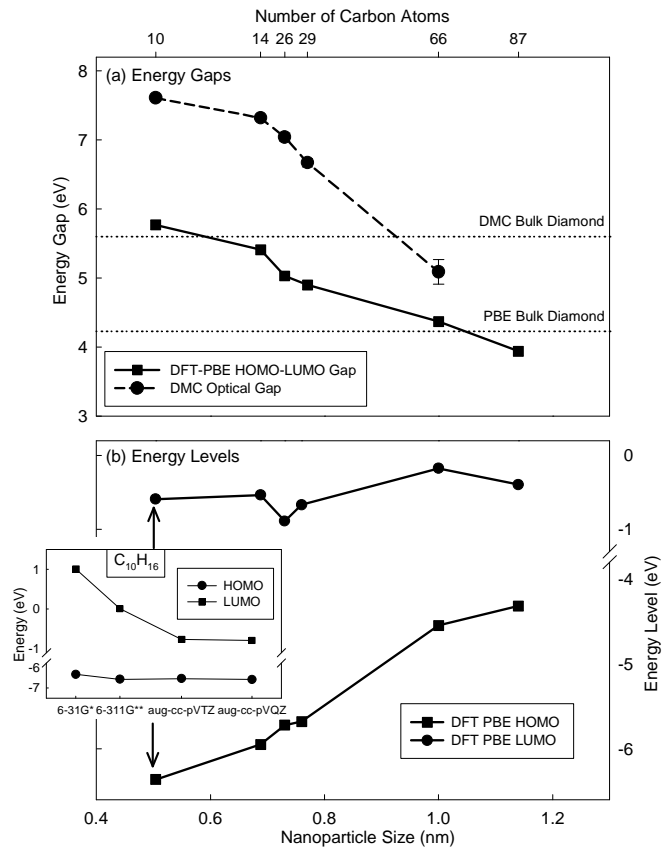}
  \caption{Calculated size dependence of (a) DFT HOMO-LUMO gap and DMC
    optical gap, and (b) DFT HOMO and LUMO eigenvalues. The inset
    shows the convergence of the DFT HOMO and LUMO eigenvalues with
    basis set.}
  \label{gap_fig}
\end{figure}

The results of our DFT calculations for the single-particle states of
the diamondoids are shown in Fig.~\ref{gap_fig} and listed in
Table~\ref{gap_table}.  Figure~\ref{gap_fig}a shows the size
dependence of the DFT eigenvalue gap between the highest occupied
molecular orbital (HOMO) and the lowest unoccupied molecular orbital
(LUMO), calculated using the plane-wave basis set described above.
All the HOMO-LUMO gaps are converged with respect to the size of the
supercell and the plane-wave basis.  The HOMO-LUMO gap decreases as
the size of the nanoparticles increases.  For nanoparticles larger
than 1\,nm the DFT-PBE HOMO-LUMO gap is smaller than the DFT-PBE gap
of bulk diamond.

Figure~\ref{gap_fig}b shows the size dependence of the individual DFT
HOMO and LUMO eigenvalues.  When the size of the nanoparticles is
reduced, the HOMO eigenvalue decreases continuously, as predicted by
standard models of quantum confinement.\cite{yoffe2002} The lower
symmetry of C$_{14}$H$_{20}$ does not alter the trend in the HOMO
eigenvalue.  In contrast, the LUMO eigenvalue displays almost no
quantum confinement and is nearly independent of size.  This behavior
differs from that of H-terminated Si and Ge
nanoparticles\cite{williamson2002}, in which both the HOMO and LUMO
exhibit clear quantum confinement.

\begin{figure}
  \includegraphics[width=\linewidth,clip=true]{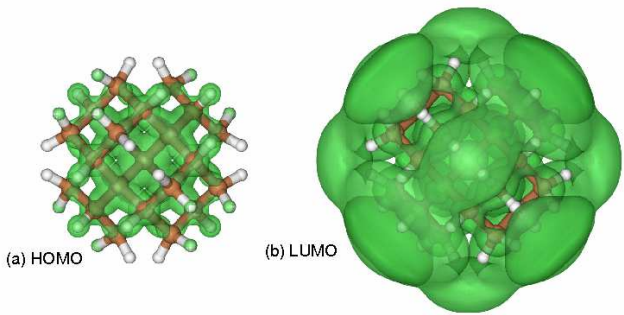}
  \caption{(Color) Isosurface plots  of the square of the (a) HOMO and
  (b) LUMO of  C$_{29}$H$_{36}$.  The green isosurfaces include 50\%
  of the charge in each orbital.}
  \label{orbital_fig}
\end{figure}

The origin of the anomalous size dependence of the LUMO energy can be
understood by comparing the HOMO and the LUMO\@.
Figure~\ref{orbital_fig} shows isosurfaces of the HOMO and LUMO of
C$_{29}$H$_{36}$.  The HOMO is localized on the C-C and C-H covalent
bonds inside the nanoparticle, while the LUMO is a delocalized state
with considerable charge outside of the H atoms terminating the
surface.  As the size of the nanoparticles increases, one expects the
HOMO to evolve smoothly into the valence-band maximum of bulk diamond.
In contrast, for larger diamondoids, the LUMO remains localized on the
surface and is closer in nature to a defect or surface level within
the gap of the bulk material; the LUMO does not evolve into the
conduction-band minimum.  This surface nature causes the optical gaps
of the larger diamondoids to lie below the bulk gap.

The surface nature of the LUMO is also responsible for some of the
previous controversies in evaluating the energy of this state.  The
inset to Fig.~\ref{gap_fig}b shows an example of the convergence of
the electronic eigenvalues in a DFT-PBE calculation of
C$_{10}$H$_{16}$ using a Gaussian basis.  It shows the energy of the
HOMO and LUMO calculated with the \textsc{G03} code~\cite{gaussian}
using 6-31G*, 6-311G**, aug-ccp-vtz,\cite{dunning89} and
aug-ccp-vqz~\cite{dunning89} basis sets.  The energy of the LUMO
decreases by almost 2\,eV as more diffuse basis functions are added,
while the HOMO energy stays approximately constant.  If a localized
basis set with insufficient flexibility to describe the diffuse
character is used, the LUMO will be artificially localized close to
the nanoparticle, increasing its kinetic energy and pushing up its
energy.  However, when the large aug-ccp-vtz~\cite{dunning89} and
aug-ccp-vqz~\cite{dunning89} basis sets are used, the Gaussian DFT-PBE
HOMO-LUMO gaps agree with the plane-wave results to within 0.01\,eV.

The QMC calculations were performed with the
\textsc{casino}~\cite{casino} code using
Slater-Jastrow~\cite{foulkes2001} trial wave functions of the form
$\Psi_T = D^\uparrow D^\downarrow \exp [J]$, where $D^\uparrow$ and
$D^\downarrow$ are Slater determinants of up- and down-spin orbitals
taken from DFT calculations and $\exp[J]$ is a Jastrow correlation
factor, which includes electron-electron and electron-ion terms
expanded in Chebyshev polynomials. The Jastrow factor was optimized
using a standard variance-minimization scheme.\cite{foulkes2001} All
the QMC optical gaps were calculated using diffusion Monte Carlo
(DMC\@).\cite{foulkes2001} To converge the DMC total energies with
respect to time step and population size, a time step of $0.02$\,a.u.\
was used and the target population was at least 640 configurations in
each calculation.  The ionic cores were represented by the same TM PBE
pseudopotentials used in the DFT calculations.  The effect on the QMC
energies of using LDA or PBE functionals to generate the DFT orbitals
was also tested.  When the PBE functional and pseudopotentials were
replaced with the LDA functional and pseudopotentials, the DFT and DMC
gaps of C$_{29}$H$_{36}$ were reduced by $0.1$\,eV and $0.3(1)$\,eV
respectively.  The DFT and DMC optical gaps are therefore relatively
insensitive to the choice of functional.

The optical gaps of the diamondoids were calculated as the
difference in the DMC energy of the ground state and an excited state.
The absorption of a photon creates an excited singlet state.  The
description of such a state requires two pairs of Slater determinants.
However, for computational simplicity, we represented the excited
state by replacing the HOMO in the spin-down Slater determinant with
the LUMO\@.  Within Ziegler's sum model for restricted Hartree-Fock
states~\cite{ziegler77}, the error incurred by adopting this
mixed-state approach is equal to half the singlet-triplet splitting.
This error is typically 0.1--0.2\,eV in group-IV nanostructures of
this size~\cite{prendergast2004}, which is small compared with the
optical gaps.  Previous QMC calculations of the optical gaps of Si
nanoparticles~\cite{williamson2002} using this approach were shown to
be in excellent agreement with \textit{GW}-Bethe-Salpeter-equation
(\textit{GW}-BSE) calculations of the true singlet excitation energies.

To investigate the sensitivity of the QMC excitation energies to the
choice of single-particle states, the excited-state calculations were
repeated with three choices for the LUMO: (i) the unoccupied LUMO from
a ground-state DFT calculation; (ii) the HOMO* occupied by the excited
electron in a DFT calculation of an excited triplet; (iii) the HOMO*
occupied by the excited electron in a DFT calculation of an excited
mixed state.  The DMC excited-state energies evaluated using (i) and
(ii) were $0.3(1)$\,eV lower than the energy evaluated using (iii),
and therefore we conclude that the excitation energy is relatively
insensitive to the choice of DFT orbital used to represent the excited
electron in the Slater determinant.

The DMC optical gaps of the diamondoids are given in Table
\ref{gap_table}, and are plotted against nanoparticle size in
Fig.~\ref{gap_fig}.  DMC calculations do not suffer from the
well-known DFT ``band-gap problem,'' as they fully describe the
interaction of the valence electrons with the electron excited into
the LUMO by the absorption of a photon, so that electron-hole
correlation is accounted for.  The DMC optical gaps are significantly
larger than the DFT HOMO-LUMO gaps, as was found for Si
nanoparticles.\cite{williamson2002}  For example, the DMC optical gap
of C$_{10}$H$_{16}$ is 7.61(2)\,eV, while the DFT-PBE gap is 5.77\,eV.
Nevertheless, the DMC results confirm the qualitative trend for the
size dependence of the gap predicted by DFT-PBE\@.  The DMC
calculations predict that quantum confinement will only be observed in
diamondoids smaller than 1\,nm.  The rapid decay of quantum
confinement in diamond nanoparticles is consistent with the small
exciton Bohr radius in diamond (1.6\,nm), compared to Si (4.9\,nm).
The optical gap of C$_{66}$H$_{64}$ is below the bulk-diamond gap, as
Raty {\em et al.}~\cite{raty2003} predicted in their DFT study.  These
DMC optical-gap calculations support the interpretation of the X-ray
absorption measurements in Ref.~\cite{raty2003}, which found no
quantum confinement effects in 4\,nm diamondoids.

To calculate the electron affinity ($EA$) and ionization potential
($IP$) of representative nanoparticles, DFT and QMC calculations were
performed for the total energy differences $EA=E_N-E_{N+1}$ and
$IP=E_{N-1}-E_{N}$, where $E_N$ is the ground-state energy of a
neutral molecule with $N$ electrons.  The trial wave functions for the
$E_{N+1}$ and $E_{N-1}$ systems were constructed by adding and
removing the orbitals corresponding to the LUMO and the HOMO from the
down-spin Slater determinant.

The DMC electron affinities and ionization energies of
C$_{10}$H$_{16}$ and C$_{29}$H$_{36}$ are given in Table
\ref{gap_table}.  Both nanoparticles have a negative electron affinity
($-0.13(2)$ and $-0.29(6)$\,eV respectively).  In contrast to the
optical gaps, where the DMC values are significantly larger than the
DFT values, the DMC electron affinities agree well with the DFT
values; for example, the DFT electron affinity of C$_{10}$H$_{16}$ is
$-0.128$\,eV\@.  The NEA suggests that coating surfaces with
diamondoids could be a simple and economic method of producing
electron emitters.

Finally, we have studied the exciton binding energy, which is defined
as the difference between the quasiparticle and optical gaps, for
small diamondoids.  For Si nanoparticles, QMC and \textit{GW}-BSE
calculations have found that the exciton binding is dramatically
enhanced at the nanoscale due to the increased overlap of the electron
and hole.\cite{benedict2003} Surprisingly, the exciton binding energy
of C$_{29}$H$_{36}$ is only $1.25(10)$\,eV compared to $2.6$\,eV for
the equivalent Si$_{29}$H$_{36}$ nanoparticle.\cite{benedict2003} The
origin of this reduced exciton binding can again be traced to the
different nature of the LUMO in C and Si nanoparticles.  The
delocalized LUMO in C$_{29}$H$_{36}$ (see Fig.~\ref{orbital_fig}b) has
a much smaller dipole matrix element with the HOMO than the equivalent
orbital in Si$_{29}$H$_{36}$, reducing the exciton binding.

In conclusion, we have performed DFT and QMC calculations of the
single-particle and optical gaps of H-terminated C nanoparticles.  Our
DFT and QMC calculations confirm the predictions of Raty \textit{et
al.}~\cite{raty2003} that quantum confinement effects disappear in
nanoparticles larger than 1\,nm.  We have shown that the LUMO of
H-terminated C nanoparticles is a delocalized surface state, in
contrast to Si and Ge nanoparticles, where the LUMO is core-confined.
The delocalized nature of the LUMO results in an anomalously small
exciton binding, a negative electron affinity, and optical gaps of
larger nanoparticles that are below the bulk gap.

We thank J.\ Grossman for performing the Gaussian DFT calculations,
and J.-Y.~Raty, T.~van Buuren, and J.~Dahl for helpful conversations.
N.D.D.\ acknowledges a fellowship in the LLNL Summer Institute on
Computational Chemistry and Materials Science.  This work was
performed under the auspices of the US Department of Energy by the
University of California, Lawrence Livermore National Laboratory under
contract No.~W-7405-Eng-48.  N.D.D.\ and R.J.N.\ acknowledge financial
support from EPSRC, UK.  Computational resources were provided by LLNL
and the Cambridge-Cranfield HPCF.

\end{document}